# Task-based Assessment of Deep Networks for Sinogram Denoising with A Transformer-based Observer

Yongyi Shi[a], Ge Wang[a], Xuanqin Mou[b]
[a]Biomedical Imaging Center, Rensselaer Polytechnic Institute, Troy, NY, USA
[b]Institute of Image Processing and Pattern Recognition, Xi'an Jiaotong University, Xi'an, Shaanxi, China

## ABSTRACT

A variety of supervise learning methods are available for low-dose CT denoising in the sinogram domain. Traditional model observers are widely employed to evaluate these methods. However, the sinogram domain evaluation remains an open challenge for deep learning-based low-dose CT denoising. Since each lesion in medical CT images corresponds to a narrow sinusoidal strip in sinogram domain, here we proposed a transformer-based model observer to evaluate sinogram domain supervised learning methods. The numerical results indicate that our transformer-based model well-approximates the Laguerre-Gauss channelized Hotelling observer (LG-CHO) for a signal-known-exactly (SKE) and background-known-statistically (BKS) task. The proposed model observer is employed to assess two classic CNN-based sinogram domain denoising methods. The results demonstrate a utility and potential of this transformer-based observer model in developing deep low-dose CT denoising methods in the sinogram domain.

## INTRODUCTION

Low-dose CT reconstruction is a hot topic in the CT field. In this application, it is crucial that the CT data noise due to a low number of x-ray photons can be well suppressed. One natural way is to preprocess sinogram data before image reconstruction, given the merit that the quantum noise is element-wise independent, and then the processed data can be fed to a reconstruction algorithm. More than a decade ago, Li et al. proposed a penalized likelihood method for quantum noise suppression as related to low-dose CT by designing a statistical model for projection data [1]. Wang et al. presented the penalized weighted least-squares method for projection domain denoising [2]. Manduca et al. [3] and Balda et al. [4] proposed two efficient algorithms for projection domain denoising, involving bilateral filtering and structural adaptive filtering. Zhang et al. proposed a projection domain denoising method based on dictionary learning [5]. Recently, convolutional neural networks (CNNs) were developed for supervised learning-based sinogram denoising [6-9]. These networks are trained by minimizing the error between low-dose and full-dose sinograms, achieving high performance in terms of image quality metrics such as root mean square error (RMSE), peak signal-to-noise ratio (PSNR), structural similarity index metric (SSIM) [10], or feature similarity index metric (FSIM) [11]. However, it is well-known that such metrics may not consider the noise correlation well, which always present in CT images. To take the strong noise correlation in CT images into account, the modulation transfer function (MTF) [12, 13] and noise power spectrum (NPS) [14] are widely used to evaluate image quality. However, Fourier-based metrics assume linearity of the imaging process, and MTF and NPS would not be proper to evaluate image quality obtained using non-linear image reconstruction methods, such as CNN-based networks.

To assess image quality from CNN-based CT reconstruction, the Bayesian ideal observer is an appropriate choice for optimizing the parameters of a CNN for signal detection tasks (e.g., detection of a lesion). Realization of a Bayesian ideal observer needs to know all statistical information of background and noise, which would be impractical in general. Alternatively, the Hotelling observer (HO) is a simplified ideal observer model that employs the Hotelling discriminant to maximize the signal-to-noise ratio of the test statistic [9]. However, the full implementation of the HO requires the estimation and inversion of a covariance matrix that can be computationally infeasible in the case of large data dimensionality [15]. Channelized Hotelling observer (CHO) approximates the HO with efficient channels, which is computationally tractable because of the reduction in data dimensionality [16]. A variety of CHOs were proposed with different channels. The CHO with Laguerre-Gauss (LG) channels is common in scenarios where the signal is known exactly and rotationally symmetric [17, 18]. A CHO with singular vectors (SV) assumes that the imaging system is linear (or at least to be decomposed into linear and nonlinear processes) and that the system response is known [19]. CHOs with partial least squares (PLS) channels overcome the limitations of the previous two models; that is, when the signal is not known exactly or not rotationally symmetric, or when the imaging system response is not known [20]. Alternatively, CHOs with Gabor channels facilitate the assessment of asymmetric text objects, do not require the imaging system response to be known, and are not constrained by the rotationally symmetric property [21]. However, all the aforementioned CHOs are still linear observers, which is suboptimal compared to the ideal observer (IO) performance for detection tasks with CT images.

Recent efforts have been primarily focused on training network-based model observes, which provide an alternative approach to CT image quality assessment. Kopp et al. proposed a CNN-based model observer in a CT liver lesion detection task [22]. Kim et al. proposed a CNN-based model observer for signal-known-exactly (SKE) and background-known-statistically (BKS) detection in breast CT images [23]. Gong et al. proposed a CNN-based model observer to predict human performance in CT tasks with various anatomical backgrounds [24]. These CNN-based model observers are employed to evaluate the traditional reconstruction methods. Since the objective evaluation of image quality for CNN-based low-dose CT reconstruction remains largely unsettled, Li et al. assessed deep network-based denoising methods with CHOs [25]. Theoretically, an ideal observer achieves the upper bound of the image quality assessment performance on the detection task. In this regard, an ideal observer that evaluates a sinogram denoise algorithm should be designed to assess the best signal detectability directly from sinogram data. By doing so, the dedicated observer would not be influenced by an image reconstruction algorithm and its parameters. Indeed, the variation in the reconstruction algorithm may affect partial volume, beam hardening, directional noise artifacts, motion artifacts, and so on. Developing a sinogram-based ideal observer involves numerous steps that bring all above-mentioned factors together. Could such a sinogram-based observer model be well established? Although existing studies suggested that it would be possible to detect a specific object from sinogram data [26], the observer model must be well designed to accurately simulate the human observer performance, which is rather complicated.

In this article, we report our preliminary study on developing the first of its kind sinogram-oriented ideal observer model; i.e., we leverage the property of sinogram data for the detection task by mimicking a CHO in the image domain formed through filtered backprojection (FBP) reconstruction. We will demonstrate the feasibility of building an observer model that takes sinogram data only. We believe that it is meaningful because sinogram data reserve the content of original information and reflect the intrinsic quality of the CT hardware and preprocessing algorithm. An ideal observer that attains the upper bound of the quality assessment performance will be used to investigate not only the CT device performance but also the performances of sinogram preprocessing and reconstruction algorithms.

Our observer model focuses on the object detection task in a local region, which discriminates a local signal from its background in reference to the recognizing capability of the human vision system. The main challenge of building the model observer is that in the sinogram domain a local image structure corresponds to a narrow sinusoidal strip and all the trips of structures are overlapped to form the entire sinogram. This property of the sinogram does not match the convolutional nature of CNNs, since they specifically extract local structural information. In recent years, inspired by the progresses in the natural language processing field (NLP), the transformer architecture [27] has been applied in the field of computer vision [28], including image quality assessment [29, 30]. The transformer architecture is applicable in the sinogram domain [31], since transformer designs a self-attentional mechanism to capture global interactions [32].

In this paper, we propose a transformer-based model observer in the sinogram domain for SKE/BKS detection, where the background is known statistically due to random background artifacts. Note that our motivation is to evaluate the influence of artifacts and noise in the supervised learning mode, where the full-dose sinogram is known exactly. Hence, the background artifacts can be obtained by subtracting the full-dose sinogram from the corresponding low-dose sinogram. We implement the transformer-based model observer in these background artifacts. The proposed transformer-based model observer is then employed to assess modern supervised CNN-based sinogram denoising methods. The canonical CNN-based denoising methods are identified for analysis, including RED-CNN [33], DnCNN [6, 34]. The remainder of the paper is organized as follows. Section II describes the background on binary signal detection task and presents the transformer-based model observer. Section III reports numerical studies and our evaluation results with respect to each of the selected denoising networks. Section IV discusses issues and makes the conclusion.

## METHODOLOGY

**Laguerre-Gauss Channelized Hotelling Observer**

To evaluate detectability, we conduct the two alternative forced choice (2-AFC) detection task. The hypotheses for signal-absent ($H_0$) and signal-present ($H_1$) are given by:

$$H_0 : \mathbf{g} = \mathbf{f}_b + \mathbf{f}_n \tag{1}$$

$$H_1 : \mathbf{g} = \mathbf{f}_s + \mathbf{f}_n \tag{2}$$

where $\mathbf{g}$ is a column vector of a given image, $\mathbf{f}_s$ is a signal-present background, $\mathbf{f}_b$ is a signal-absent background, $\mathbf{f}_n$ is noise. Our goal is to assess the quality of CT images in the SKE/BKS detection task, where the background is known statistically, which is defined by random background artifacts.

For assessment of CT image quality, the HO provides the upper bound of the detection performance among all linear model observers. To avoid estimating a large covariance matrix needed in the HO, the CHO approximates the HO performance with efficient channels of much-reduced dimensionality. In the CHO, an image $\mathbf{g}$ is transformed to $\mathbf{v}$ by

$$\mathbf{v} = \mathbf{U}\mathbf{g} \quad (3)$$

where $\mathbf{U}$ is the channel matrix. LG-CHO uses a LG function for the channel matrix and can approximate the CHO performance well for a rotationally symmetric signal in a known location. The LG function is defined as follows:

$$U_j(r|a) = \frac{\sqrt{2}}{a} exp\left(\frac{-\pi r^2}{a^2}\right) L_j\left(\frac{2\pi r^2}{a^2}\right) \quad (4)$$

with the Laguerre polynomial function:

$$L_j(x) = \sum_{i=0}^{j}(-1)^i \binom{j}{i}\frac{x^i}{i!} \quad (5)$$

where $r$ is the distance between a point of interest and the center, $j$ denotes the $j$th channel, and $a$ is the width of the Gaussian function. In this study, we use 5 channels with 2 mm width of the Gaussian function. Figure 3 shows the appearances of the 5 channels under given Gaussian widths.

The CHO template estimated from training images is computed by:

$$\mathbf{w} = \Delta \mathbf{v}^T \mathbf{K}_v^{-1} \quad (6)$$

where $\Delta \mathbf{v}$ is the mean difference between the signal-present and signal-absent transformed images, and $\mathbf{K}_v^{-1}$ is the image covariance matrix, $T$ denotes the transpose. The CHO decision variable can be computed by

$$d = \mathbf{w}^T \mathbf{v} \quad (7)$$

In this study, the LG-CHO is applied to ROIs cropped from a reconstructed residual image around lesion centers. The circular ROIs are of size 64 pixels × 64 pixels, where the lesion size was of 9 pixels. We extract 1,000 signal-present and 1,000 signal-absent samples to be our training set.

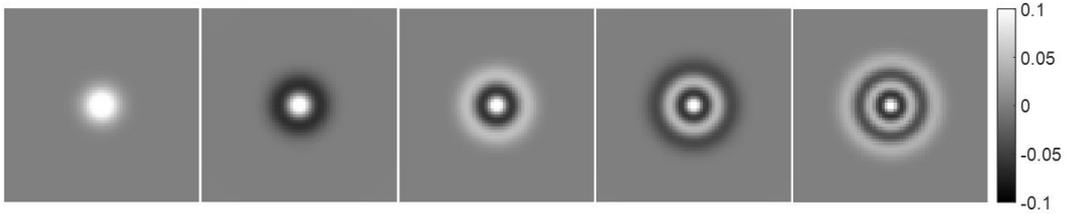

Figure 1. Appearance of the five LG channels of size 64 pixels × 64 pixels and Gaussian width of 10 pixels along both axes.

**Low-dose CT data preparation**

We used the full-dose CT slice of 3 mm thickness from a conventional clinical dataset used in "the 2016 NIH-AAPM-Mayo Clinic Low Dose CT Grand Challenge" prepared by Mayo Clinic to generate the dataset for this study [31]. Figure 2 shows the flowchart for generating a signal-present image in liver region. Specifically, we inserted a circularly shaped signal near the center of each region of interest (ROI) with an elevated CT value by 20 Hounsfield Unit (HU) to present a challenge of low contrast lesion detection. After forward projection, we obtained the full-dose signal-present sinogram, where the signal is a narrow sinusoidal strip in the sinogram domain. Poisson noise was superimposed to the sinogram to produce the low-dose signal-present sinogram. The filtered backprojection (FBP) algorithm was then employed for image reconstruction. Table I lists the simulation parameters for fan-beam CT reconstruction. The background artifacts were obtained by subtracting the full-dose signal-absent sinogram/image from the low-dose signal-present sinogram/image. After then, ROIs of 64 pixels × 64 pixels were extracted from the image to evaluate the LG-CHO method. In the sinogram domain, sinusoidal strips of 64 bins × 1160 views were extracted and reshaped to a matrix as the signal-present data for analysis by the transformer-based model observer.

Table I. Simulation parameters.

| Parameters | Value |
|---|---|
| Source to Iso-Center Distance | 570 mm |
| Minimum angle | -0.456 |
| Maximum angle | 0.4526 |
| The number of detector bins | 672 |
| The number of views | 1160 |
| Reconstructed pixel size | 0.8 mm × 0.8 mm |
| Reconstructed matrix size | 512 pixels × 512 pixels |

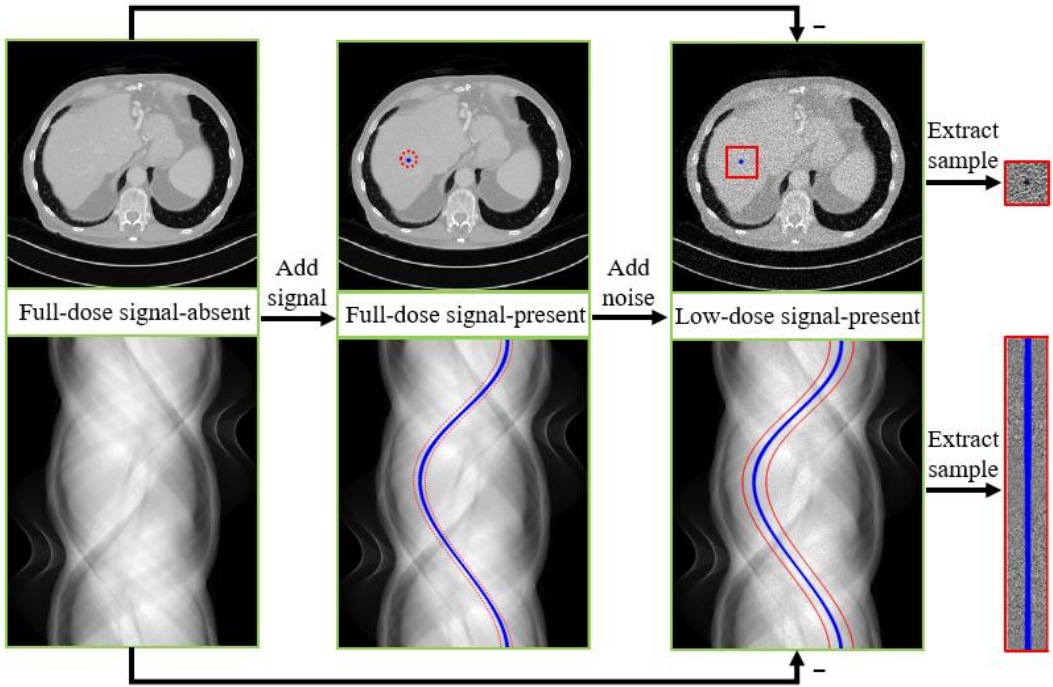

Figure 2. The flowchart for data generation.

**Transformer-based model observer**

The transformer architecture is featured by a global attention mechanism, and thus can be used to extract sinogram features from different view angles, outperforming the CNN-based approach. In traditional vision transformer, image is cut into patches as input tokens to capture their spatial relationship in nature images. Since each sinogram view represents one measurement of an imaging target, modeling relations between these views can help the network learn correlation among different views [26]. Thus, we split each sinogram view as an input token. Figure 3 shows the top-level structure of the proposed transformer-based model observer. We modeled the detection task as a classification problem. For the 2-AFC detection task, the randomly selected signal-present and signal-absent sinogram pair serve as the input to the transformer. Note that $\dot{F}'\in\mathbb{R}^{H\times D}/F'\in\mathbb{R}^{H\times D}$ randomly represents the signal-present/signal-absent or signal-absent/signal-present sinogram pairs, where $H$ and $D$ are the number of detector bins and views respectively. As often used in the vision transformer, an extra embedding $F_0$ is appended to the beginning of the input sinogram. we add the learnable position encodings $P\in\mathbb{R}^{(H+1)\times D}$ to the corresponding $F\in\mathbb{R}^{(H+1)\times D}$ to keep the position information. The calculation of the encoder is formulated as

$$\begin{aligned} y_0 &= [F_0 + P_0, F_1 + P_1, \dots, F_N + P_N], \\ q_i &= k_i = v_i = y_{i-1}, \\ y_i' &= LN(MHA(q_i, k_i, v_i) + y_{i-1}), \\ y_{i-1} &= LN(MHA(y_i') + y_i'), \quad i = 1, \dots, L \\ [E_0, E_1, \dots, E_N] &= y_L \end{aligned} \quad (8)$$

where L represents the number of the encoder-decoder layers, and the output of the encoder $E\in\mathbb{R}^{(H+1)\times D}$ has the same size to the input $F$.

The decoder takes another sinogram with an extra embedding as the input. The output of the encoder, $E$, is used as an input of the decoder in the second MHA layer. The calculation of the decoder can be formulated as

$$\begin{aligned} y_L &= [E_0, E_1, \dots, E_N], \\ z_0 &= [\dot{F}_0 + \dot{P}_0, \dot{F}_1 + \dot{P}_1, \dots, \dot{F}_N + P_N], \\ q_i &= k_i = v_i = z_{i-1}, \\ z_i' &= LN(MHA(q_i, k_i, v_i) + z_{i-1}), \\ y_{i-1} &= LN(MHA(y_i') + y_i'), \quad i = 1, \dots, L \\ [\dot{E}_0, \dot{E}_1, \dots, \dot{E}_N] &= y_L \end{aligned} \quad (9)$$

The output $\dot{E}\in\mathbb{R}^{(H+1)\times D}$ of the decoder is finally obtained to feed into the following MLP head.

The MLP head consists of two fully-connected (FC) layers, and the first FC layer is followed by the ReLU activation. The second FC layer is followed by the Sigmoid activation to predict the labels of the sinogram pair.

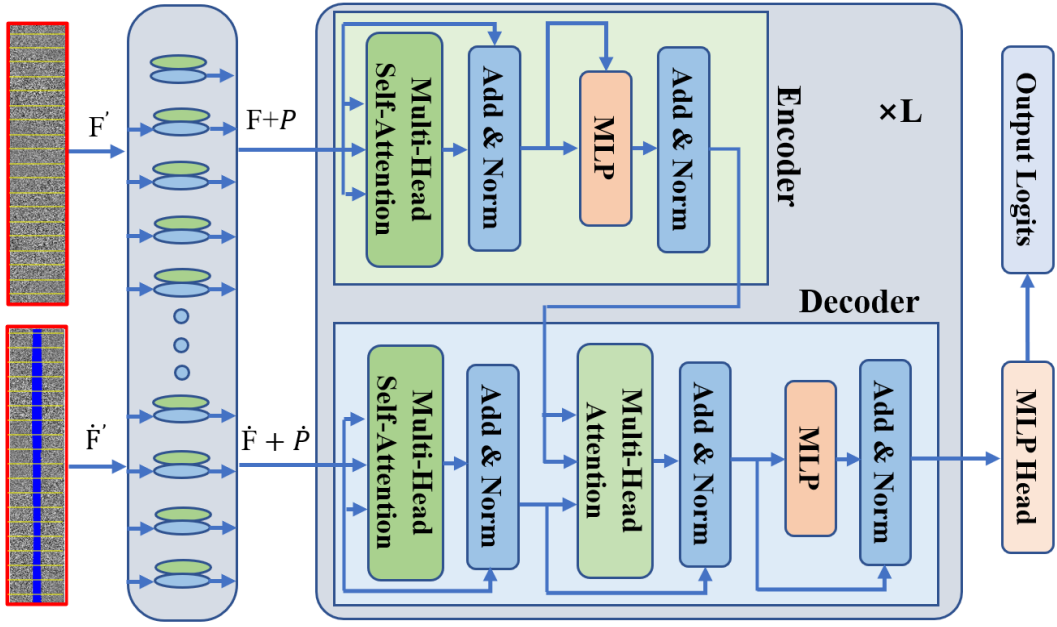

Figure 3. The architecture of the proposed transformer-based model observer.

During network training, the hyper-parameters were set as follows. The number of encoder and decoder layers was set to 1. The number of heads in the MHA was set to 4. The transformer dimension was set to 1,160. The dimension of the MLP in the encoder and decoder was set to 1,024. The dimension of the first FC layer in the MLP head was 128. The dimension of the patches fed into the transformer was 64 × 1160. The training process was done using the Adam optimizer with a batch size of 16. The initial learning rate was 0.00005, with the cosine learning rate decay. The entropy cross loss was used for training. The number of epochs was set to 200. Our network was implemented using Pytorch framework with a single NVIDIA GeForce 3090 GPU. To train our network, we used 5,000 signal-present sinograms and 5,000 signal-absent sinograms for the 2-AFC task. We divided the generated datasets by a 19:1 ratio for training and validation datasets.

**Denoising networks**

To evaluate the proposed model observer applied to deep learning-based low-dose CT algorithms, the two classical networks were employed for projection domain denoising: RED-CNN and DnCNN. A real clinical dataset shared by Mayo Clinics for "the 2016 NIH-AAPM-Mayo Clinic Low Dose CT Grad Challenge" was used to validate the performance of the networks. In the training stage, 9 patients were selected. In the testing stage. 1 patient was selected. We used the normal-dose images from the dataset to simulate low dose sinograms. For RED-CNN and DnCNN, the network architectures were made the same as that in [28] and [29] respectively. The training process was conducted using the Adam optimizer with a batch size of 16. The initial learning rate was 1e-5 for RED-CNN and 1e-3 for DnCNN. The mean-square-error (MSE) loss function was used. The number of epochs was set to 100.

**Evaluation of the detection performance**

We used the percent correct ($P_c$) as a detection performance measure for the model observer, which is defined as

$$P_c = \frac{1}{N_t}\sum_{i=1}^{N_t} o_i \tag{10}$$

where $N_t$ is the number of test trails and $o_i$ is binary decision variables, where $o_i$ can only be 1 or 0, corresponding to correct or incorrect detection results respectively. To obtain $o_i$, we compare the correct answer with the logit of the network output. In the case of LG-CHO, we select the image which has the largest decision variable between 2 input images and compare it with the correct answer. We use 1,000 sinogram pairs (i.e., 1,000 signal absent sinograms and 1,000 signal present sinograms) as the test dataset to evaluate the performance of our proposed transformer-based model observer. The corresponding 1,000 image pairs (i.e., 1,000 signal absent images and 1,000 signal present images) were used as the test dataset to evaluate the performance of LG-CHO.

## RESULTS

**Detectability in different noise levels**

Figure 4 shows sinograms/images with 100, 50, 25 thousand incident photons. With the decrease of incident photons, the signal-to-noise ratio (SNR) of the sinograms/images is gradually reduced. Some fine details are increasingly obscured by noise, which are detrimental in clinical diagnosis; e. g., lesion detection. We used the numerical observer models to evaluate the detectability by extracting ROIs in the sinograms/images.

Figure 5 shows some samples in the training dataset. The circular signal was inserted in the center of each ROI in the artifact images to simulate a challenge of low contrast lesion detection. The signal in sinogram corresponds a narrow strip, and is difficult to distinguish visually in both artifact images and sinograms because of strong noise.

Figure 6 shows the performance curves of the signal detectability versus the dose level. It can be seen from the curves that the performance trends among the LG-CHO and transformer-based model observer are similar. This simiarity shows that our proposed transformer-based model observer can be used to evaluate the signal detection task in the sinogram domain.

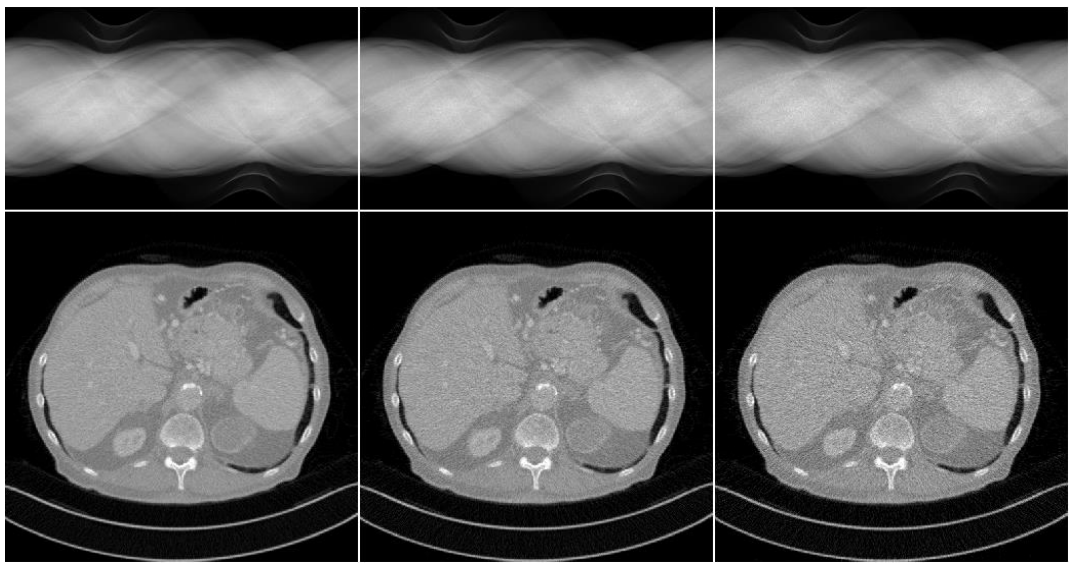

Figure 4. From left to right are the sinograms/reconstructed images with 100, 50, 25 thousand incident photons per detector element. The sinograms are shown with a display window [0, 9], while the corresponding images are in a display window [0, 0.035] mm$^{-1}$.

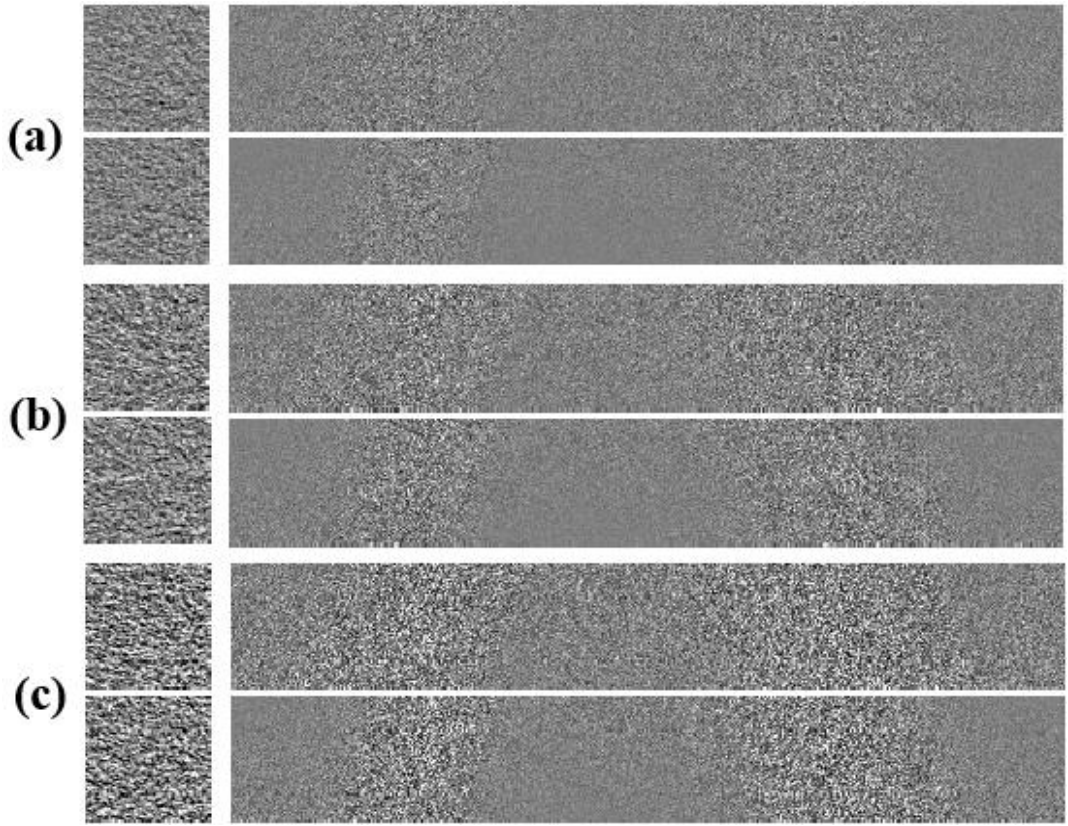

Figure 5. Representative samples in the training dataset. From (a)-(c) are the images/sinograms with 100, 50, 25 thousand incident photons per detector element. In each group of (a)-(c), the top is the signal-present images/sinograms, bottom is the signal-absent images/sinograms. The left is the ROIs from the images in a display window [-0.005, 0.005], while the right is the ROIs from the sinograms in a display window [-0.5, 0.5].

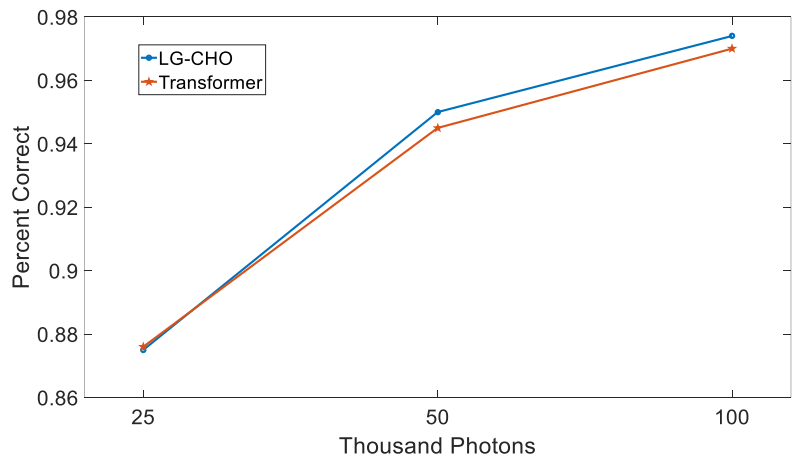

Figure 6. The lesion detectability curves with respect to the dose level.

**Evaluation of the Denoising networks**

Figure 7 shows the sinograms/images with 25 thousand incident photons, using RED-CNN and DnCNN respectively. The noise was suppressed by the denoising networks. Table II shows the quantitative results using different denoising networks on the whole testing set. It can be seen that the

RED-CNN method produced the images with the highest peak signal-to-noise ratio (PSNR), root-mean-square error (RMSE) and structure similarity index (SSIM). DnCNN also improved on these quantitative metrics. However, some fine details were blurred. For example, the dots indicated by the red arrow can be seen in the original low-dose CT image but were blurred by the denoising networks. This is because both RED-CNN and DnCNN employ the MES loss to train the network, which does not guarantee the task-specific optimality; e. g., lesion detection. Especially for DnCNN, the introduced secondary artifacts clearly degraded the image quality.

Figure 8 shows typical samples in the training dataset. After processed by either RED-CNN or DnCNN, the noise was suppressed. However, some features were somehow distorted or blurred in the residual sinogram. Both blurred details in the image domain and stretched structures in the residual projection domain compromised the performance of the numerical observer models.

Figure 9 shows the performance curves of the signal detectability associated with different methods. It can be seen that the performance trends for the LG-CHO and transformer-based model observer are in excellent agreement. Interestingly, the proposed transformer-based model observer performed better than LG-CHO. However, the detectablitity decreased after the image was prepossed by the denoising networks, which is consistent to the visual inspection. The fine details were removed by the denosing networks, which reduced the detectability. Hence, it is desireable to propose a method that can improve the metrics of RMSE/PNSR/SSIM and the LG-CHO/Transformer-based observer simultaneously.

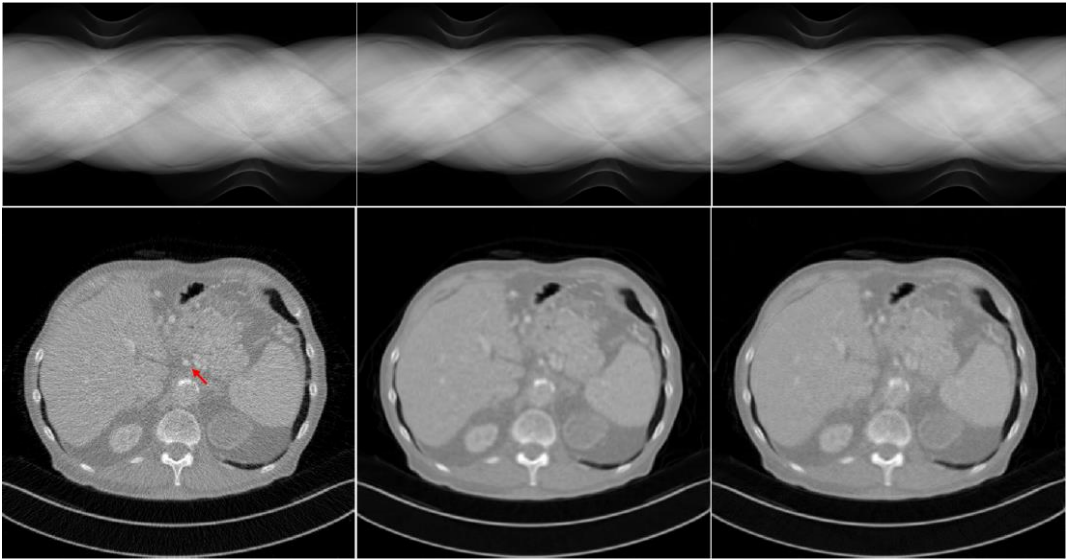

Figure 7. From left to right are the sinograms/images with 25 thousand incident photons using RED-CNN and DnCNN respectively. The top is the sinograms in a display window [0, 9], while the bottom is the corresponding images in a display window [0, 0.035] mm$^{-1}$.

Table II. Quantitative results from the different denoising networks.

| Methods | Low dose | RED-CNN | Dn-CNN |
|---|---|---|---|
| RMSE | 0.0069 | 0.0025 | 0.0038 |
| PNSR | 43.4594 | 52.0860 | 48.3616 |
| SSIM | 0.9608 | 0.9981 | 0.9879 |

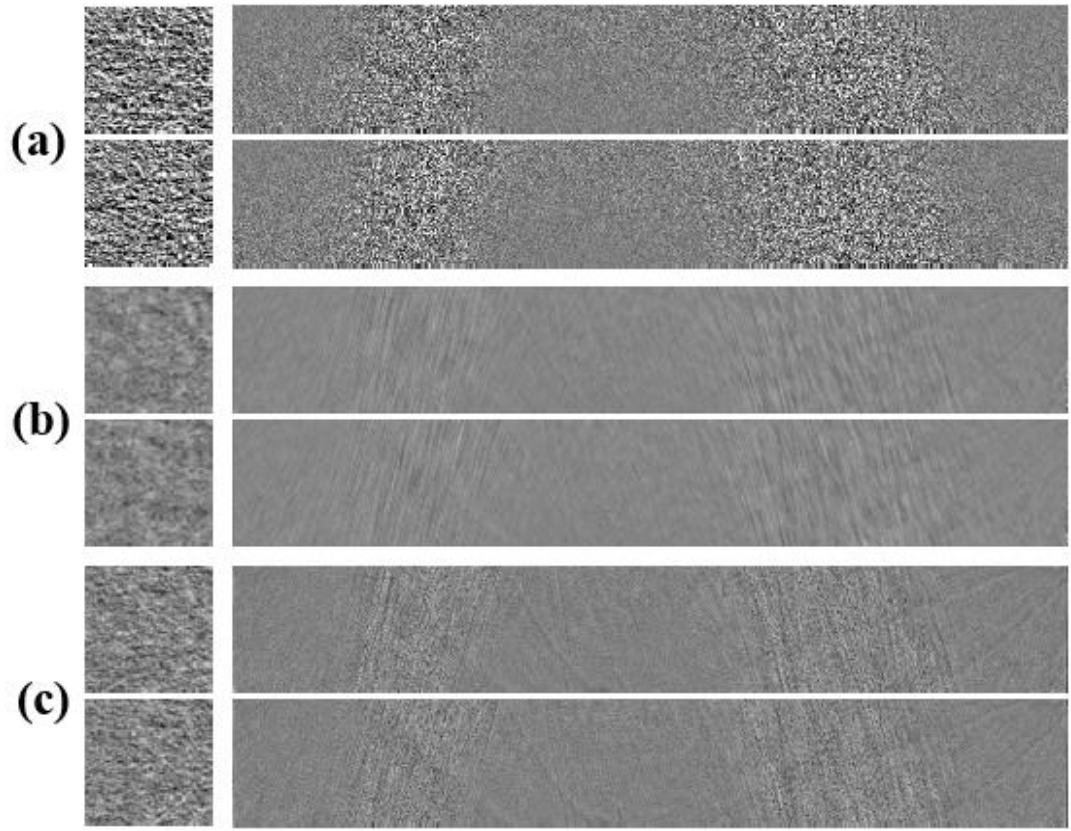

Figure 8. Typical samples in the training dataset. (a)-(c) The images/sinograms with 25 thousand incident photons per detector element, with RED-CNN and DnCNN respectively. For each group of (a)-(c), the top is the signal-present images/sinograms, while the bottom is the signal-absent images/sinograms. The left shows the ROIs from the images in a display window [-0.005, 0.005], while the right is the ROIs from the sinograms in a display window [-0.5, 0.5].

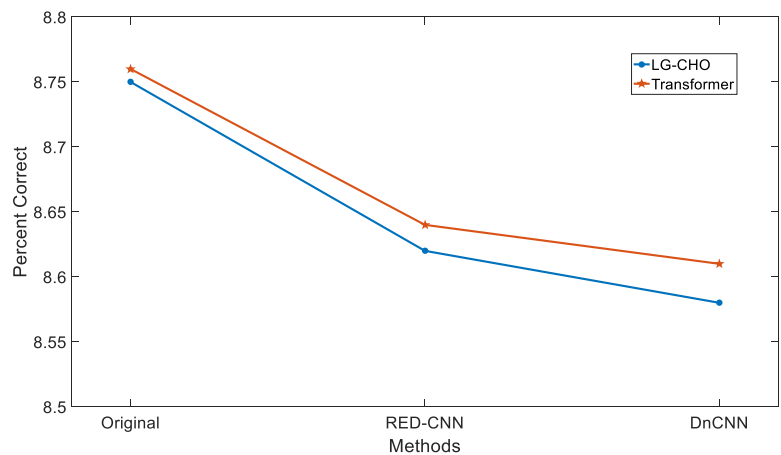

Figure 9. The lesion detectability curves with respect to the dose level.

## DISCUSSIONS AND CONCLUSION

In this study, we have demonstrated that the transformer-based model observer yields a performance similar to the LG-CHO for detection of a circular signal under an SKE/BKS condition regarding a residual background. Hence, the reference images are required to calculate the residual background. However, in practice the reference images are usually difficult to obtain. Furthermore, the proposed transformer-based method targets the ideal observer, which is still not the same as the human observer. Implementing an anthropomorphic model observer using a transformer-based model is another interesting topic, which will not rely on reference images and bring the results closer to what the human observer achieves. A large amount of data labeled by a human observer should be prepared for training the transformer-based anthropomorphic model observer, which is a future topic for our research.

We have only evaluated two low-dose CT sinogram domain denoising methods. However, there are numerous sinogram denoising methods including both traditional methods and deep learning-based methods. In our future work, we will evaluate more denoising methods, including traditional methods. Furthermore, the impact of the parameters of the denoising networks on task-specific performance measures needs to be evaluated; e. g., the depth of the networks.

In conclusion, we have proposed a transformer-based model observer to evaluate low-dose CT sinogram domain supervised learning methods. With the transformer, the proposed observer model has yielded a performance similar to that of the LG-CHO model. However, LG-CHO has a strict assumption (i.e., a rotationally symmetric signal, a known location, and a stationary background) to approximate the ideal linear observer performance. In contrast, the transformer-based observer model is less restrictive. Additionally, the evaluation in the sinogram domain can avoid the image reconstruction process, which suggests future avenues to develop and optimize sinogram domain supervised learning methods. In general, the further work is to build an ideal observer model based on sinogram data so that we could approach the upper bound of the evaluation performance of sinogram-based low-dose CT denoising.

## REFERENCES

[1] T. Li, X. Li, J. Wang et al., "Nonlinear sinogram smoothing for low-dose X-ray CT," *IEEE Transactions on Nuclear Science*, vol. 51, no. 5, pp. 2505–2513, 2004.

[2] J. Wang, T. Li, H. Lu et al., "Penalized weighted least-squares approach to sinogram noise reduction and image reconstruction for low-dose X-ray computed tomography," *IEEE Transactions on Medical Imaging*, vol. 25, no. 10, pp. 1272–1283, 2006.

[3] A. Manduca, L.F. Yu, J.D. Trzasko et al., "Projection space denoising with bilateral filtering and CT noise modeling for dose reduction in CT," *Medical Physics*, vol. 36, no. 11, pp. 4911–4919, 2009.

[4] M. Balda, J. Hornegger, B. Heismann, "Ray contribution masks for structure adaptive sinogram filtering," *IEEE Transactions on Medical Imaging*, vol. 31, no. 6, pp. 1228–1239, 2012.

[5] H. Zhang, L. Zhang, Y. Sun, J. Zhang, "Projection domain denoising method based on dictionary learning for low-dose CT image reconstruction," *Journal of X-ray Science and Technology*, vol. 23, no. 5, pp. 567-78, 2015.

[6] M. U. Ghani, W. C. Karl, "CNN based sinogram denoising for low-dose CT," *Mathematics in Imaging*, 2018.

[7] M. Meng, S. Li, L. Yao et al., "Semi-supervised learned sinogram restoration network for low-dose CT image reconstruction," *SPIE Medical Imaging*, p. 11312, 2020.

[8] Y. J. Ma, Y. Ren, P. Feng et al., "Sinogram denoising via attention residual dense convolutional neural network for low-dose computed tomography," *Nuclear Science and Techniques*, vol. 32, no. 4, pp. 1-14, 2021.

[9] L. Chao, P. Zhang, Y. Wang et al., "Dual-domain attention-guided convolutional neural network for low-dose cone-beam computed tomography reconstruction," *Knowledge-Based Systems*, p. 109295, 2022.

[10] Z. Wang, A. C. Bovik, H. R. Sheikh et al., "Image quality assessment: from error visibility to structural similarity," *IEEE transactions on image processing*, vol. 13, no. 4, pp. 600-612, 2004.

[11] L. Zhang, L. Zhang, X. Mou et al., "FSIM: A feature similarity index for image quality assessment," *IEEE transactions on Image Processing*, vol. 20, no. 8, pp. 2378-2386, 2011.

[12] S. Richard, D. B. Husarik, G. Yadava et al., "Towards task‐based assessment of CT performance: system and object MTF across different reconstruction algorithms," *Medical Physics*, vol. 39, no. 1, pp. 4115-4122, 2012.

[13] S. N. Friedman, G. S. Fung, J. H. Siewerdsen et al., "A simple approach to measure computed tomography (CT) modulation transfer function (MTF) and noise‐power spectrum (NPS) using the American College of Radiology (ACR) accreditation phantom," *Medical Physics*, vol. 40, no. 5, p. 051907, 2013.

[14] H. H. Barrett, K. J. Myers, "Foundations of image science," *John Wiley & Sons*, 2013.

[15] H. H. Barrett, K. J. Myers, B. D. Gallas et al., "Megalopinakophobia: its symptoms and cures," *SPIE Medical Imaging*, pp. 299-307, 2001.

[16] B. D. Gallas, H. H. Barrett, "Validating the use of channels to estimate the ideal linear observer," *JOSA A*, vol. 20, no. 9, pp. 1725-1738, 2003.

[17] X. He, S. Park, "Model observers in medical imaging research," *Theranostics*, vol. 3, no. 10, p. 774, 2013.

[18] R. Zeng, S. Park, P. Bakic et al., "Evaluating the sensitivity of the optimization of acquisition geometry to the choice of reconstruction algorithm in digital breast tomosynthesis through a simulation study" *Physics in Medicine & Biology*, vol. 60, no. 3, p. 1259, 2015.

[19] S. Park, J. M. Witten, K. J. Myers, "Singular vectors of a linear imaging system as efficient channels for the Bayesian ideal observer," *IEEE transactions on medical imaging*, vol. 28, no. 5, pp. 657-668, 2008.

[20] J. M. Witten, S. Park, K. J. Myers, "Partial least squares: a method to estimate efficient channels for the ideal observers," *IEEE Transactions on medical imaging*, vol. 29, no. 4, pp. 1050-1058, 2010.

[21] D. Gomez-Cardona, C. P. Favazza, S. Leng et al., "Task-specific efficient channel selection and bias management for Gabor function channelized Hotelling observer model for the assessment of x-ray angiography system performance," *Medical physics*, vol. 48, no. 7, pp. 3638-3653, 2021.

[22] F. K. Kopp, M. Catalano, D. Pfeiffer et al., "CNN as model observer in a liver lesion detection task for x-ray computed tomography: A phantom study," *Medical physics*, vol. 45, no. 10, pp. 4439-4447, 2018.

[23] G. Kim, M. Han, H. Shim et al., "A convolutional neural network‐based model observer for breast CT images," *Medical physics*, vol. 47, no. 4, pp. 1619-1632, 2020.

[24] H. Gong, J. G. Fletcher, J. P. Heiken et al., "Deep-learning model observer for a low-contrast hepatic metastases localization task in computed tomography," *Medical physics*, vol. 49, no. 1, pp. 70-83, 2022.

[25] K. Li, W. Zhou, H. Li et al., "Assessing the impact of deep neural network-based image denoising on binary signal detection tasks," *IEEE transactions on medical imaging*, vol. 40, no. 9, pp. 2295-2305, 2021.

[26] Q. De Man, E. Haneda, B. Claus et al., "A two-dimensional feasibility study of deep learning-based feature detection and characterization directly from CT sinograms," *Medical physics*, vol. 46, no. 12, pp. e790-e800, 2019.

[27] A. Vaswani, N. Shazeer, N. Parmar et al., "Attention is all you need," *Advances in neural information processing systems(NIPS)*, 2017.

[28] A. Dosovitskiy, L. Beyer, A. Kolesnikov et al., "An image is worth 16x16 words: Transformers for image recognition at scale," arXiv preprint arXiv:2010.11929, 2020.

[29] J. You, J. Korhonen, "Transformer for image quality assessment," *IEEE International Conference on Image Processing (ICIP)*, pp. 1389-1393, 2021.

[30] M. Cheon, S. J. Yoon, B. Kang et al., "Perceptual image quality assessment with transformers," *Proceedings of the IEEE/CVF Conference on Computer Vision and Pattern Recognition*, pp. 433-442, 2021.

[31] L. Yang, Z. Li, R. Ge et al., "Low-Dose CT Denoising via Sinogram Inner-Structure Transformer,". *IEEE Transactions on Medical Imaging*, Early Access, 2022.

[32] J. Liang, J. Cao, G. Sun et al., "Swinir: Image restoration using swin transformer," *Proceedings of the IEEE/CVF International Conference on Computer Vision*, pp. 1833-1844, 2021.

[33] H. Chen, Y. Zhang, M. K. Kalra et al., "Low-dose CT with a residual encoder-decoder convolutional neural network," *IEEE transactions on medical imaging*, vol. 36, no. 12, pp. 2524-2535, 2017.

[34] K. Zhang, W. Zuo, Y. Chen et al., "Beyond a gaussian denoiser: Residual learning of deep CNN for image denoising," *IEEE transactions on image processing*, vol. 26, no. 7, pp. 3142-3155, 2017.